\documentclass{IEEEtran4PSCC}
\ifCLASSINFOpdf
   \usepackage[pdftex]{graphicx}
\else
   \usepackage[dvips]{graphicx}
\fi
 
\usepackage{amssymb}
\usepackage{amsthm}
\usepackage{amsmath}
\usepackage{xcolor}
\usepackage{comment}
\usepackage{nomencl}
\usepackage{ifthen}
\usepackage{etoolbox}
\usepackage{bbm}
\usepackage{algorithm}
\usepackage{enumerate}
\usepackage{float}
\usepackage{booktabs}
\usepackage{tikz}
\usepackage{soul}
\usepackage{multirow}
\usepackage{algpseudocode}
\usepackage{graphicx}
\usepackage{epstopdf}
\usepackage{dsfont}

\definecolor{JHUblue}{RGB}{0, 45, 114}
\definecolor{JHUwhite}{RGB}{255, 255, 255}  
\definecolor{JHUblack}{RGB}{ 0, 0, 0}  
\definecolor{color1}{RGB}{215, 38, 56}
\definecolor{color2}{RGB}{63, 136, 197}
\definecolor{amber}{RGB}{244, 157, 55}
\definecolor{color4}{RGB}{62, 58, 83}
\definecolor{plot1}{RGB}{31, 119, 180}
\definecolor{plot2}{RGB}{255, 127, 14}
\definecolor{plot3}{RGB}{44, 160, 44}
\definecolor{amber}{rgb}{1.0, 0.49, 0.0}

\DeclareMathSymbol{\Minus}{\mathbin}{AMSa}{"39}

\allowdisplaybreaks

\makenomenclature
\renewcommand{\nomgroup}[1]{%
  \item[%
    \ifthenelse{\equal{#1}{A}}{A. \textit{Sets}}{}%
    \ifthenelse{\equal{#1}{C}}{B. \textit{Parameters}}{}%
    \ifthenelse{\equal{#1}{B}}{C. \textit{Variables}}{}
    ]%
    \vspace{10pt}\hspace*{-\leftmargin}\vspace{10pt}%
}

\theoremstyle{plain}

\makeatletter
\let\old@ps@headings\ps@headings
\let\old@ps@IEEEtitlepagestyle\ps@IEEEtitlepagestyle
\def\psccfooter#1{%
    \def\ps@headings{%
        \old@ps@headings%
        \def\@oddfoot{\strut\hfill#1\hfill\strut}%
        \def\@evenfoot{\strut\hfill#1\hfill\strut}%
    }%
    \def\ps@IEEEtitlepagestyle{%
        \old@ps@IEEEtitlepagestyle%
        \def\@oddfoot{\strut\hfill#1\hfill\strut}%
        \def\@evenfoot{\strut\hfill#1\hfill\strut}%
    }%
    \ps@headings%
}
\makeatother

\psccfooter{%
        \parbox{\textwidth}{\hrulefill \\ \small{24th Power Systems Computation Conference} \hfill \begin{minipage}{0.2\textwidth}\centering \vspace*{4pt} \includegraphics[scale=0.06]{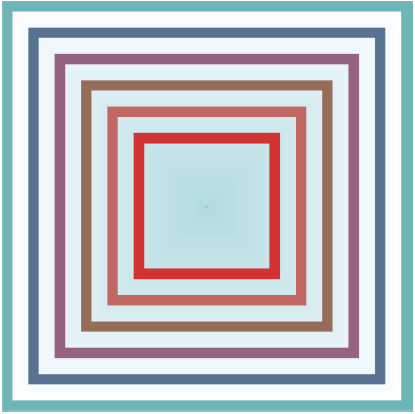}\\\small{PSCC 2026} \end{minipage} \hfill \small{Limassol, Cyprus --- June 8-12, 2026}}%
}

\begin{document}
%
\title{Learning Reachability of Energy Storage Arbitrage}

\author{
\IEEEauthorblockN{Tomás Tapia\IEEEauthorrefmark{2}\\ Agustin Castellano\IEEEauthorrefmark{2}}
\IEEEauthorblockA{\IEEEauthorrefmark{2}Electrical and Computer Engineering Department \\
Johns Hopkins University\\
Baltimore, Maryland, United States\\
\{ttapia1, acaste11\}@jh.edu}
\and
\IEEEauthorblockN{Enrique Mallada\IEEEauthorrefmark{2}\\ Yury Dvorkin\IEEEauthorrefmark{2}\IEEEauthorrefmark{1}}
\IEEEauthorblockA{\IEEEauthorrefmark{1}Civil and Systems Engineering Department \\
Johns Hopkins University\\
Baltimore, Maryland, United States\\
\{mallada, ydvorki1\}@jh.edu}
}
\maketitle

\begin{abstract}
Power systems face increasing weather-driven variability and, therefore, increasingly rely on flexible but energy-limited storage resources. Energy storage can buffer this variability, but its value depends on intertemporal decisions under uncertain prices. Without accounting for the future reliability value of stored energy, batteries may act myopically, discharging too early or failing to preserve reserves during critical hours. This paper introduces a stopping-time reward that, together with a state-of-charge (SoC) range target penalty, aligns arbitrage incentives with system reliability by rewarding storage that maintains sufficient SoC before critical hours. We formulate the problem as an online optimization with a chance-constrained terminal SoC and embed it in an end-to-end (E2E) learning framework, jointly training the price predictor and control policy. The proposed design enhances reachability of target SoC ranges, improves profit under volatile conditions, and reduces its standard deviation.
\end{abstract}



\printnomenclature

\section{Introduction} \label{Sec1:Introduction}
\subsection{Motivation and Scope}
Power systems experience increased weather-driven variability that impacts both renewable energy generation and demand patterns. This variability also impacts wholesale electricity markets, where day-ahead (DA) and real-time (RT) market outcomes increasingly depart from each other \cite{bienstock2024risk}, which in turn causes system operators and market participants to adjust their decision-making routines. Efficient and reliable operations under these conditions require anticipating scarcity and preserving flexibility before weather-driven stressors materialize. Energy storage systems, such as batteries, are well positioned to alleviate this variability due to their fast response capability. However, translating this technical capability into reliable system outcomes is non-trivial, because the economic value of stored energy is time-dependent and hinges on the ability to account for \emph{intertemporal} linkages between storage decisions at different time periods. Without estimating the intertemporal economic value of storage under uncertain and variable wholesale prices, these resources may respond to changing RT conditions myopically, leading to suboptimal economic and reliability performance. That is, in terms of energy arbitrage, battery assets may be over-discharging on early price spikes or over-charging when flexibility is scarce, leading to scarcity precisely at the time periods when the system most needs reserves. We refer to these periods of time as the system's \emph{critical hours}.

System operators have therefore explored targeted mechanisms to maintain appropriate state-of-charge (SoC) levels to ensure energy availability during critical hours. These mechanisms aim to ensure compliance with a given minimum SoC limit or a sustained rate of charging or discharging, so storage can support system reliability when operating conditions tighten. At the grid level, the California Independent System Operator (CAISO) proposed a Minimum State of Charge (MSOC) requirement that sets a minimum SoC constraint for DA and RT scheduling so multi-hour batteries remain available during peak-demand and evening ramp periods \cite{CAISO2024}. In the Electric Reliability Council of Texas (ERCOT), NPRR 1186 seeks to enhance SoC awareness and telemetry, improving accounting in system operations, and establishing obligations for limited-energy storage resources so that system operators can estimate deliverability during critical hours \cite{PUCT2024}. At the distribution level in California, PG\&E’s Seasonal Aggregation of Versatile Energy (SAVE) program operates a virtual power plant, where the utility issues week-ahead, hour-by-hour capacity needs signals to demand-side aggregators for aligning residential loads, batteries, and solar panels with local reliability needs \cite{PGE_SAVE_VPP_2025}. However, prescribing a minimum SoC level directly, or enforcing administrative rules outside of market clearing, can introduce efficiency losses and implementation complexity.

Two challenges arise when attempting to internalize reliability to the power system through market signals to an individual energy storage asset. First, \emph{reachability}: does there exist a sequence of control actions (charge, discharge, idle) for the battery that can reach and maintain a desired SoC level or range by time $\tau$ associated with the critical hours, for at least a minimum fraction of uncertainty realizations, while satisfying operational limits? Consequently, we want to avoid the \emph{point of not return} where satisfying the minimum fraction of uncertainty realizations is not possible even with maximum control of the battery. Determining reachability and simultaneously optimizing profit further compound complexity. Second, \emph{probability quantification}: given a reward/penalty signal to the battery, what is the probability that the battery’s SoC will lie within the target SoC range during critical hours? Answering this question analytically requires solving a corresponding chance constraint, which is often expensive or intractable under realistic, heavy-tailed price dynamics. Even scenario-based surrogates can become computationally burdensome for multi-period problems.
\vspace{-2mm}
\begin{figure}[H]
    \centering
    \includegraphics[width=1.04\linewidth, trim={4.6cm 19.1cm 6.5cm 4.2cm},clip]{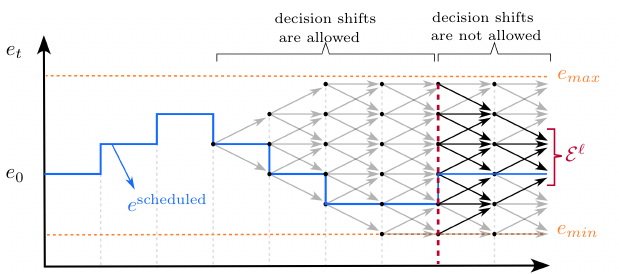}
    \caption{State-of-charge and action trajectories for the battery arbitrage problem, where the blue line denotes the scheduled energy trajectory (e.g., from a  previous schedule).}
    \label{fig:enter-label2}
    \vspace{-1mm}
\end{figure}
Figure \ref{fig:enter-label2} shows the two challenges. The blue line shows the scheduled SoC $e_t$ reaching time $\tau$, after which decision shifts (branching tree) adapt charge/discharge/idle actions to realized prices, subject to $[e^{\min}, e^{\max}]$ limits. \emph{Reachability} corresponds to whether some trajectory can enter and maintain the SoC target range $\mathcal{E}^{\ell}$. The point of no return is the latest time before $\tau^*$ at which, even under full battery control, no trajectory can reach $\mathcal{E}^{\ell}$. Beyond $\tau^*$, decision shifts are not allowed, so feasibility must already be secured. \emph{Probability quantification} then concerns the likelihood that the terminal SoC lies within $\mathcal{E}^{\ell}$ during the critical hours, i.e., the fraction of trajectories that could lead and remain within the SoC target range.

This paper addresses the system’s lack of preparedness for potential critical hours by embedding a \textit{stopping-time} reward in the battery asset objective. The stopping-time and SoC target band constitute a new attribute that the battery asset could provide to the system for reliability purposes. Thus, the stopping-time will work as a time barrier that triggers before the point of no return, switching the policy from arbitrage to preservation mode to ensure the target SoC. The reward series encourages the battery to enter a reserve-preserving mode sufficiently early, providing the operator with a margin to prepare and operate the system to avoid unnecessary system stress. Economically, these rewards act as time-varying price premia for energy retained before critical hours, aligning (private) arbitrage incentives with the systemic (public) objective of resilience. Methodologically, we formulate the arbitrage problem as an online optimization that enforces reliability through a \emph{chance-constraint-style} terminal SoC target, penalizing trajectories that fail to enter a prescribed SoC range. To align incentives, forecasts, and battery dispatch, we adopt a task-based end-to-end (E2E) approach in which the energy price predictor (represented as an uncertainty set) is trained jointly with a downstream optimization task to maximize arbitrage profit and stopping-time reward. Rather than minimizing a generic forecast error, the price predictor is optimized to produce decisions that meet the desired coverage level of the energy prices and calibrate the predictor to maximize the arbitrage and stopping reward profit.

\subsection{Literature review}
In the context of energy storage systems, computing the future value of stored energy is one of the major challenges for an energy arbitrageur, i.e., an entity that buys energy at a low price and sells it when the price is higher. Recent work on estimating the value of stored energy falls into two strands: empirical valuation of price‑taking arbitrage using historical prices and data \cite{sioshansi2009estimating}, and analytical valuation of battery assets under multi‑stage price uncertainty \cite{xu2020operational}.

Previous work has also examined storage profitability and arbitrage as an \textit{inventory optimization} problem. On siting, sizing, and technology choice, \cite{dvorkin2016ensuring} studies the profitability of battery storage under high penetration levels of renewable energy, \cite{zhang2021arbitrage} proposes a strategic framework that evaluates technologies under differing efficiencies and life-cycle costs, and \cite{singhal2020pricing} introduces alternative SoC market models and analyzes arbitrage within different price settings. On market design and operations, \cite{krishnamurthy2017energy} evaluates grid-scale arbitrage in the DA market with corrective actions in RT, while \cite{zheng2022arbitraging} develops an analytical stochastic dynamic programming approach for variable-efficiency storage arbitrage in RT. Integration with system operations has been explored via chance-constrained pricing models for storage participation \cite{qi2023chance,qi2025locational}, cooperative DA dispatch where battery storage could provide multiple regulation services and improve reliability in \cite{zhang2023day}, and nested Markov Decision Process (MDP) formulations that co-optimize arbitrage and frequency regulation in \cite{cheng2016co}.

The battery arbitrage problem has also been analyzed as an online optimization problem, since energy prices are revealed sequentially, and charge/discharge decisions must be made irrevocably under SoC constraints. In the context of online problems, \textit{stopping-time criteria} arise from optimal stopping theory and from competitive analysis in online algorithms. Some common examples are the secretary problem \cite{babaioff2008online,degroot1968some} and a broader theory of stopping rules \cite{degroot1968some,pichler2022risk}. A stopping rule is a policy to determine the stopping-time $\tau$ with respect to the observed process based only on information revealed up to time $t$, not on future outcomes. Recent work studies time-consistency of risk measures and its relation with optimal stopping criteria \cite{pichler2022risk,shapiro2012time,bielecki2025time}. Building on these principles, \cite{lorenz2009optimal} develops $k$-search algorithms for online optimization that yield buy/sell thresholds with worst-case guarantees, and \cite{zhang2011online} extends the analysis to selling $k$ units over $n$ periods in a multi-unit setting.

From a control-policy and reinforcement learning (RL) perspective, energy storage operations under arbitrage can be framed to enforce limits and penalize undesired states. On the control side, \cite{harsha2014optimal} studies optimal storage management using a discrete-time average-cost stochastic dual programming, deriving dual-threshold and greedy policies. In model-free RL, Q-learning has been used for energy arbitrage \cite{wang2018energy, castellano2020learning}, and deep Q-learning in \cite{cao2020deep}. Safety considerations have been incorporated through primal–dual methods that enforce constraints during learning \cite{paternain2022safe,ding2020natural}, as well as constrained policy optimization approaches that provide additional safety guarantees \cite{achiam2017constrained,yang2020projection}. In the control context, \textit{reachability analysis} is a method that has been studied to provide safety guarantees for systems with unknown or uncertain dynamics to reach a predetermined goal. A survey of reachability and set-propagation techniques that balance tightness and scalability across linear, nonlinear, and hybrid systems is presented in \cite{althoff2021set}, and practical deployments for autonomous vehicles are presented in \cite{althoff2010reachability}. Data-driven variants estimate reachable sets from simulation \cite{castellano2023learning} and attach statistical guarantees using holdout calibration \cite{dietrich2025data}, adversarial sampling and random-set modeling \cite{lew2021sampling}, and decomposition of deterministic inputs from stochastic disturbances to obtain high-probability sets \cite{jafarpour2025probabilistic}. Complementary learning-based approaches provide distribution-free, finite-sample valid prediction sets for chance-constraints using conformal methods \cite{zhao2024conformal}, and decision-focused training of uncertainty sets tailored to robust optimization, improving out-of-sample decision quality \cite{wang2023learning}.

Recent machine learning (ML) approaches to inventory-like control under price uncertainty advance along several fronts. Data-driven forecasting is integrated directly into decision pipelines \cite{lee2024online}, and learned thresholds guide operational choices \cite{sang2022electricity}. Switching costs and action frictions are modeled via threshold-based online policies in \cite{lechowicz2023online,li2020online}. Beyond decoupled predict-then-optimize workflows, end-to-end (E2E) learning jointly trains forecasting and control to improve task performance in \cite{donti2017task}, while conformal calibration further enhances uncertainty quantification and decision robustness within E2E frameworks in \cite{yeh2024end}. Existing ML-driven approaches could be extended to infer the impact of exogenous factors on prices and ensure safe SoC trajectories given a specific downstream task. However, none of these methods directly internalize the stopping-time decision, i.e., when to adjust the chosen trajectory to ensure sufficient SoC during critical hours.

Our approach introduces a stopping-time reward that internalizes the operator’s reliability objective, where batteries are rewarded only if they transition into and maintain a reserve-preserving mode through the end of the period. This makes it possible to  reconcile the system and asset perspectives by endogenizing the value and likelihood of the terminal SoC reachability to both system reliability and asset profitability. This design minimizes incentives to take arbitrage actions during the online optimization problem, promotes no changes in the SoC trajectory prior to critical hours, and ensures that compensation is tied to assured performance rather than ad hoc constraints.

\subsection{Contributions}
The goal of this paper is to bring reachability analysis into inventory optimization used for energy storage performing arbitrage and endogenously optimize the stopping-time to assure, with a user-defined probability level, a feasible future trajectory to reach a given SoC at the end of the planning horizon. Specific contributions include:
\begin{itemize}
    \item We extend the battery arbitrage problem, formulated as an online decision process, by including a stopping-time reward and a terminal SoC penalty that internalize the operator’s preference to preserve energy ahead of critical hours. The stopping-time reward steers battery assets toward feasible SoC trajectories and the reachability of the system operator’s SoC targets.
    \item We embed the stopping-time reward and a terminal SoC penalty in an end-to-end (E2E) framework for battery arbitrage to mitigate the profit impact of price volatility under exogenous uncertainty. The E2E approach provides robustness guarantees by calibrating the energy-price predictor to a given downstream optimization task, resulting in increased profit under worst-case price realizations while keeping a user-defined likelihood of meeting the terminal SoC target band.
    \item We provide an empirical comparison of the E2E approach against two baselines, a reward/penalty Deep Q-learning model and a mixed-integer stochastic model, capturing different levels of complexity. Using real day-ahead price data, we highlight the trade-off between reliability provision and the asset's profit. Results show that time-indexed control signals, implemented as stopping-time rewards, steer batteries toward reliability-consistent behavior and offer a market-compatible alternative to minimum SoC requirements.
\end{itemize}


\section{Energy arbitrage under uncertain price with stopping reward}


Consider the battery arbitrage problem, where a battery asset seeks to maximize its profit by charging ($c_t$) and buying energy when prices are low, and discharging ($d_t$) and selling energy when prices are high for $t$ in time-horizon $\mathcal{T}\!=\!\{1,\dots,T\}$. Under uncertain energy prices, the main challenge for battery assets is to quantify the value of stored energy and decide how to manage the state-of-charge ($e_t$) before observing the realization of the (still) uncertain energy price $(\boldsymbol{\lambda}_t)$ to maximize its profit. From the perspective of the battery asset, the arbitrage problem can be formulated in an inventory optimization manner as follows:
\begin{subequations} \label{P0}
\begin{align}
    \max_{e,c,d}~~&\mathbb{E}_{\boldsymbol{\lambda}} \Big[\sum_{t=1}^T \boldsymbol{\lambda}_t (d_t - c_t) \Big] \label{P0:objective_function}\\
    \text{s.t.}~& e_{t=0} ~= e_0, \label{P0:initial_state}\\
    \Big\{&e^{\rm min} \leq e_t \leq e^{\rm max} && \label{P0:storage_capacity}\\
    &e_t ~= \hat{\eta} e_{t-1} - \frac{d_t}{\eta_d} + c_t \eta_c, && \label{P0:battery_dynamics}\\
    &0 \leq d_t \leq P, \quad 0  \leq c_t \leq P \Big\}, && \forall \; t \in \mathcal{T}, \label{P0:charge_discharge_limit}\\
    & \mathbb{P}_{\boldsymbol{\lambda}}\{ e_{\tau} \in \mathcal{E}^{\ell} \} \; \geq 1 - \epsilon, && \forall \; \tau \in \mathcal{T^{\rm target}}, \label{P0:chance_constraint}
\end{align}    
\end{subequations}
where \eqref{P0:objective_function} is the objective function that maximizes the expected arbitrage profit under uncertain energy prices. Degradation mechanisms and associated costs are not considered. However, they can be incorporated using linear approximation models \cite{maov_battery2013} without altering the problem's complexity. Eq. \eqref{P0:initial_state} and \eqref{P0:storage_capacity} set the initial SoC and the storage capacity limits within $[e^{\min}, e^{\max}]$. Eq. \eqref{P0:battery_dynamics} represents the SoC dynamics, where $\eta_c$ and $\eta_d$ are the charging and discharging efficiencies, and $\hat{\eta}$ is a given self-discharge factor. Eqs. in \eqref{P0:charge_discharge_limit} impose the charging and discharging limits at the rate $P$ for each period. Charging and discharging actions are constrained to be mutually exclusive. Finally, \eqref{P0:chance_constraint} specifies a chance constraint under price uncertainty, requiring the terminal SoC to lie within the target band $\mathcal{E}^{\ell} = [\ell^{\min}, \ell^{\max}]$, not necessarily returning to $e_0$ (note that we require that $e_0\in \mathcal{E}^{\ell}$). The lower bound $\ell^{\min}$ can be used to support peak-demand supply, while the upper bound $\ell^{\max}$ can be used to relieve congestion. This constraint must hold with probability at least $1-\epsilon$, where $\epsilon$ is a user-defined risk tolerance level, over all pre-defined critical hours, i.e., $\forall \tau \in \mathcal{T}^{\rm target}$. For simplicity, we will consider $\mathcal{T}^{\rm target} = \left\{T\right\}$.

Battery asset actions are driven by the information and comprehensive risk assessment of the uncertain energy price sequence $\{\lambda_t\}_{t=1}^T$, as well as by the risk tolerance of the decision-maker ($\epsilon$). In order to incentivize an early transition into the \emph{reserve-preserving} mode, i.e., when charging or discharging is suboptimal relative to the current time period but will be beneficial at a later time period, we introduce an additional term, the stopping-time reward series $\{\hat{r}_t\}_{t=1}^T$. The purpose of this reward is to steer batteries into this mode, so the system operator has sufficient resources to serve load during later critical hours and to take corrective actions if the \emph{reachability} of the chance-constraint \eqref{P0:chance_constraint} is not guaranteed.

For simplicity, consider normalized charge ($c_t$), discharge ($d_t$), and idle ($g_t$) actions based on their maximum rate $P$. Thus, the stochastic arbitrage problem considering stopping-time rewards can be expressed as follows:
\begin{subequations} \label{P1}
    \begin{align}
        \max_{c,d,e,z,g}~&\mathbb{E}_{\boldsymbol{\lambda}} \Big[\sum_{t=1}^{T} \boldsymbol{\lambda}_t P (d_t - c_t) + \sum_{t=1}^{T} \hat{r}_t z_{t} \Big] \label{P1:objective_function}\\
        \text{s.t.}\quad& e_{t=0} = e_0, ~ z_{t=0} = 0, \label{P1:initial_state}\\
        \Big\{& \eqref{P0:storage_capacity} - \eqref{P0:battery_dynamics},~\eqref{P0:chance_constraint}, \nonumber\\
        & z_t \geq z_{t-1}, \label{P1:stop_monotonicity}\\
        & z_t \geq 1 - (d_t + c_t + g_t), \label{P1:stop_bound_1}\\
        & c_t \leq 1 - z_t, \quad d_t \leq 1 - z_t, \label{P1:stop_bound_2}\\
        & 0 \leq d_t+c_t+g_t \leq 1, \label{P1:charge_discharge_idle_limits}\\
        & c_t,~d_t,~g_t \in [0,1], \label{P1:charge_discharge_idle_def}\\
        & z_t \in \{0,1\} \Big\}, \qquad \qquad \forall ~t\in\mathcal{T}, \label{P1:stopping_variable_def}
    \end{align}
\end{subequations}
where $z_t$ is the binary variable that takes the value $1$ if the battery asset enters the reserve-preserving mode, and $g_t$ corresponds to the idle action that works as a slack variable that allows the battery asset to delay the decision of stopping. Parameter $\hat{r}_t$ is a stopping-time reward sequence determined by either an asset owner or jointly by the asset owner and the system operator, and, in principle, is set ex ante and independent of the realization of $\lambda_t$. Eq. \eqref{P1:objective_function} is the objective function that includes the reward term. Eq. \eqref{P1:stop_monotonicity} guarantees monotonicity of the stopping decision, i.e., that it cannot be undone at a later time stage. Eqs. \eqref{P1:stop_bound_1} and \eqref{P1:stop_bound_2} correspond to the stopping decision logic, where no action can be taken if the stopping variable $z_t = 1$. Eq. \eqref{P1:charge_discharge_idle_limits} relates the charging, discharging, and idle actions. 

Note that, due to the monotonicity of $z_t$ in time $t$, a simple modification of the objective function guarantees that the reward is collected exactly once. Replace the last term in \eqref{P1:objective_function} with $\sum_{t=1}^{T} \hat{r}_t \bigl(z_t\!-\! z_{t-1}\bigr)$ which awards $\hat{r}_\tau$ only at the stopping-time $\tau$ when $z$ transitions from $0$ to $1$. This helps interpretability and, in practice, pays exactly once at the switch of $z$, avoiding double counting. We use this in Section B.

\subsection{Sample Averaged Approximation (SAA) formulation}
SAA is a standard technique for solving stochastic optimization problems by replacing a complex or intractable uncertainty representation with empirical averages, typically computed with a mixture of i.i.d. simulated or historical scenarios. In this case, we enforce the chance-constraint in \eqref{P0:chance_constraint} using a mixed-integer reformulation with scenario indicators $w_k$. We select SAA over chance-constrained reformulations to better fit empirical data without imposing restrictive distributional assumptions. Relative to the chance-constrained reformulations, SAA incurs (i) additional computational complexity that scales with the number of scenarios, and (ii) data inefficiency under naive sampling, especially in the context of rare or high-impact events that drive risk or constraint violations. This problem could be tackled using \emph{importance sampling} by sampling scenarios from a proposal distribution $q(\xi)$ that over-represents informative regions (e.g., peaks, ramps), and re-weighting each sample by its likelihood ratio $w(\xi)=p(\xi)/q(\xi)$, where $p$ is the target distribution. Using SAA with continuous decision variables could improve computational performance; however, it makes the learning step more challenging (e.g., because the decision-tree surrogate relies on a discrete action space).

In our setting, SAA serves as a benchmark example for identifying the stopping-time, i.e., the time $\tau$ at which the battery starts the reserve-preserving mode, allowing us to study how the chosen $\tau$ trades off arbitrage profit and empirical reliability and isolate its effect from the impacts of the E2E approach. Consider $\mathcal{K} = \{1,\ldots,K\}$, a set of historical scenarios; thus, the SAA formulation for the arbitrage problem with the stopping-time reward is expressed as follows:
\begin{subequations}
    \begin{align} \label{P2}
        \max_{c,d,e,z,g, u} \quad& \frac{1}{|K|} \sum_{k=1}^{K} \sum_{t=1}^{T} \boldsymbol{\lambda}_{tk} P (d_{t} - c_{t}) + \sum_{t=1}^{T} \hat{r}_t z_{t}\\
        \text{s.t.}\quad& \eqref{P1:initial_state}-\eqref{P1:charge_discharge_idle_def} \nonumber\\
        & w_k \ell^{\rm min} + (1 - w_k) e^{\rm min} \leq e_{T}, \label{P2:lower_bound_stopping}\\
        & e_{T} \leq w_k \ell^{\rm max} + (1 - w_k) e^{\rm max}, \label{P2:upper_bound_stopping}\\
        & \frac{1}{|K|} \sum_{k=1}^K w_k \geq 1-\epsilon, \label{P2:minimum_scenarios}\\
        & z_{t},~w_{k} \in \{0,1\}, \qquad \quad \forall ~t\in\mathcal{T},~k\in\mathcal{K}, \label{P2:stopping_scenario_variable_def}
    \end{align}
    \label{eq:saa}
\end{subequations}
where constraints \eqref{P2:lower_bound_stopping}-\eqref{P2:minimum_scenarios} represent the sample-based mixed-integer reformulation for the chance-constraint in \eqref{P0:chance_constraint}. For a given scenario $k\in\mathcal{K}$, the inequalities \eqref{P2:lower_bound_stopping}-\eqref{P2:upper_bound_stopping} establish that the SoC at time $T$ must be within the safe bound $\mathcal{E}^{\ell} = [\ell^{\rm min},\ell^{\rm max}]$ if the binary variable $w_k = 1$. Finally, Eq. \eqref{P2:minimum_scenarios} ensures that the chance constraint is met with a user-defined probability.

\subsection{Deep Q-Learning (DQN) formulation}
The SAA formulation in \eqref{eq:saa}, however, suffers from increased computational requirements for a large number of scenarios. As an alternative to SAA, Q-learning is a well-known, model-free RL method that is suitable for small and discrete state-action spaces. Q-learning trains an action-value functional relationship $Q(s,a)$ via temporal-difference updates of the states $(s)$ and actions $(a)$. Deep Q-Learning (DQN) replaces the action-value function with a neural network (NN) approximator $Q_\theta(s,a)$ and trains it based on a predefined policy. This extension enables policies to scale to high-dimensional observations while retaining discrete actions.

In the battery arbitrage problem, DQN is well suited because the problem is inherently sequential, partially observed, and path-dependent. As a model-free algorithm, DQN learns directly from historical or simulated trajectories without requiring an explicit price model (but still requires defining the battery dynamics). It scales to high-dimensional observations via neural function approximation, and operational limits can be enforced via action masking, constraint enforcement, and violation penalties, while the stopping-time rewards and terminal penalties can be used to align the learned policy with given SoC range targets, yielding reliability-consistent behavior.

\subsubsection{Markov decision process (MDP) formulation}
Consider a finite-horizon MDP with horizon $T$. The state at time $t$ is $s_t= \big(e_t, u_t, p_t, t\big)\in\mathcal{S}$, where $e_t\in\{0,\Bar{P},\dots,\lfloor e^{\max}\rfloor\!-\!\bar{P}\}$ is the discretized SoC level, $u_t \in \{0,1\}$ is the stopping indicator, $p_t = b_t(\lambda_t)\in\{0,1,\dots,B\!-\!1\}$ is a discretized price bin, and $t\in\{1,\dots,T\}$. The charging and discharging rates are discretized by $\bar{P} \in (0, P]$, and using smaller rates will increase computation times. In practice, for example, certain storage technologies, including long-duration storage, exhibit near-constant charging and discharging rates \cite{schleifer2025exploring}. The action set is $ \mathcal{A}=\{ a_t: -1~(\text{discharge}),~0~(\text{idle}),~1~(\text{charge}),~2~(\text{stop})\}$.
Let the price $\lambda_t$ be binned using the function $b(\lambda_t)$ as follows:
\begin{subequations}
\begin{align}
    b(\lambda_t) \; & =\; \max \Bigg(0, \min\Bigg(\left\lfloor\dfrac{\lambda_t-\lambda^{\min}}{\Delta \lambda}\right\rfloor, B\!-\!1\Bigg)\Bigg), \label{clip_1}\\
    \Delta \lambda &=\dfrac{\lambda^{\max}-\lambda^{\min}}{B} \label{clip_2},
\end{align}
\end{subequations}
where the price $\lambda_t$ denotes the realized energy price at $t$ that is assigned to a bin $b(\lambda_t)$ based on a maximum number of bins $B$ and the price range $[\lambda^{\min},\lambda^{\max}]$. For the training task, $\lambda_{t+1} \in [\lambda^{\min}_{t+1}, \lambda^{\max}_{t+1}]$ is sampled i.i.d. and $p_{t+1}=b(\lambda_{t+1})$.

\subsubsection{SoC dynamics}
Let $e_{t+1}$ evolve deterministically given an SoC $(e_t)$ and an action $(a_t)$, the next SoC $(e_{t+1})$ is defined as follows: 
\begin{equation}
    e_{t+1} \;=\;
    \begin{cases}
    e_t+\bar{P}, & a_t=1,\ e_t<S-\bar{P},\\[3pt]
    e_t-\bar{P}, & a_t=-1,\ e_t>0,\\[3pt]
    e_t,   &  a_t\notin \{-1,1\},
    \end{cases}
    ~ S = \lfloor e^{\max}\rfloor \label{dqn_state}
\end{equation}
\subsubsection{Rewards}
Let us define the stopping indicator $u_t$ that is $\text{true}$ when the stopping action is made (with $u_{t=0}=\text{false}$):
\begin{equation}
    u_{t+1}=
    \begin{cases}
    u_t, & a_t \neq 2,\\
    \text{true}, & a_t = 2,
    \end{cases}
    \quad \text{and } u_{t}=\text{false}, \label{dqn_reward}
\end{equation}
Let $\lambda_t$ denote the realized price at $t$, then, when $u_t = \text{false}$, the per-step reward $\hat{r}_t$ corresponds to:
\begin{equation}
    \hat{r}_t \;=\;
\begin{cases}
\phantom{-}\lambda_t  \bar{P}, & a_t=-1,\ e_t>0,\\
0, & a_t=0,\\
-\lambda_t \bar{P}, & a_t=1,\ e_t<S-\bar{P},\\
r_{t}, & a_t=2,
\end{cases}
\end{equation}
where $\{r_t\}_{t=1}^T$ is the stopping-time reward sequence determined by the system operator, which is activated only when the decision is made and changes $u_t$ to $\text{true}$. If $u_t=\text{true}$, for the following times, $\tau \in \{t+1, T\}$, the per-step reward is set to $\hat{r}_{\tau} = 0$.
At the terminal time $T$, we add a constraint penalty and the accumulated reserve-preserving reward:
\begin{equation}
    \hat{r}_{T}\ \leftarrow\ \hat{r}_{T}
    \;-\;
    \begin{cases}
    \rho(||e_t\!-\!\ell^{\rm max}||^2 + ||\ell^{\rm min}\!-\!e_t||^2), & e_{T+1}\notin \mathcal{E}^{\ell},\\
    \ \ \ 0, & e_{T+1}\in \mathcal{E}^{\ell},
    \end{cases} \label{dqn_reserve_preserving}
\end{equation}
where $\mathcal{E}^{\ell}$ is the acceptable terminal SoC band, and $\rho>0$ is the violation penalty factor. Note that the stopping-time reward is an implementation approximation of \eqref{P1:objective_function}, where the battery owner only collects the reward at time $t$ when the stopping action is made.

\subsubsection{Policy class and Q-network}
DQN approximates the action–value function with a neural network. In our setup, we use a Multi-Layer Perceptron (MLP) with ReLU activations after the first two linear layers:
\begin{equation}
Q_\theta:\ \mathbb{R}^{3}\!\to\!\mathbb{R}^{|\mathcal{A}|}, \qquad
Q_\theta(s) \;=\; \mathrm{MLP}_{64 \rightarrow 64 \rightarrow |\mathcal{A}|}(s),
\end{equation}
where the output dimension equals the number of discrete actions $|\mathcal{A}|$. The policy is $\varepsilon$-greedy with respect to $Q_\theta$. Let the $\varepsilon$-greedy policy induced by $Q_\theta$ be the following:
\begin{equation}
\pi_\theta(s)\;=\;\arg\max_{a\in\mathcal{A}} Q_\theta(s)_a.
\end{equation}
Then, actions are selected based on the probability $\varepsilon$:
\begin{equation}
a_t \sim
\begin{cases}
\mathrm{Uniform}(\mathcal{A}_t), & \text{w.p. } \varepsilon,\\[2pt]
\arg\max_{a\in\mathcal{A}_t} Q_\theta(s_t)_a, & \text{w.p. } 1-\varepsilon,
\end{cases}
\end{equation}
which balances exploration and exploitation during training, and $\mathcal{A}_t$ corresponds only to feasible actions for $e_t$.

\subsubsection{Loss and update}
DQN approximates the action–value function $Q^\pi(s,a)$ with a neural network $Q_\theta(s,a)$ and minimizes the mean-squared error with respect to the Bellman target using a discount factor $\gamma\in(0,1)$ as follows:
\begin{equation}
\mathcal{L}(\theta)\;=\;\mathbb{E}\Big[\big(r_t+\gamma \max_{a'} Q_{\bar\theta}(s_{t+1},a')-Q_\theta(s_t,a_t)\big)^2\Big] \label{Bellman_error},
\end{equation}
where $Q_{\bar\theta}$ is a \emph{target network} whose parameters  are updated to track $Q_\theta$, and $Q_{\theta}$ is updated by stochastic gradient descent on \eqref{Bellman_error} using mini-batches sampled from a replay buffer, i.e., user defined memory that stores past transitions. 

\subsubsection{Policy extraction}
After training, for any state $s=(e,u,p,t)$, the learned policy is $\pi_\theta(s)=\arg\max_{a}Q_\theta(s)_a$,
and can be tabulated over $e \in\{0,\dots,S-1\}$ and $p\in\{0,\dots,B-1\}$ for each $t \in \mathcal{T}$.

Thus, DQN (i) interacts with a simulated or historical price environment to collect transitions $(s_t,a_t,\hat{r}_t,s_{t+1})$; (ii) stores them in the replay buffer; (iii) samples mini-batches to update $\theta$ by stochastic gradient descent; and (iv) selects actions via $\varepsilon$-greedy (or another pre-defined policy).

Given the MDP formulation and an accurate model of the storage dynamics, we do not use model-based techniques like Stochastic (Dual) Dynamic Programming (SDDP) \cite{pereira1991multi, porteiro2018towards} or Model Predictive Control (MPC) \cite{garcia1989model, castellano2025data}. We refrain from using these methods because they require specifying a stochastic model for price dynamics (or a prior over future price distributions) which is undesirable in the heavy-tailed regimes of interest. Moreover, incorporating chance constraints or probabilistic reachability targets typically entails risk-averse reformulations, which significantly increase their computational complexity and tractability.

\subsection{End-to-End (E2E) framework with a Stopping Criterion}
The battery arbitrage problem aligns well with an end-to-end (E2E) framework \cite{donti2017task}, \cite{yeh2024end}, in which a price prediction model is linked to a downstream differentiable arbitrage optimization problem and trained on the downstream objective, i.e., the prediction is made aware of its performance in a given downstream task. The predictor outputs an energy price calibrated uncertainty set, and the dispatch layer solves a robust arbitrage problem under the calibrated set. The task loss is computed from the optimizer’s outcome and then back-propagated to the predictor model, so the prediction is tuned for the quality of the decision rather than a proxy error metric.

In our battery arbitrage setting, a differentiable dispatch layer is embedded beneath a price predictor so that the parameters are updated to maximize the realized arbitrage profit in addition to the stopping-time reward. Specifically, we incorporate a stopping-time reward $\{r_t\}_{t=1}^{T}$ that compensates for an early transition into a reserve-preserving mode at time $\tau$, shaping the SoC trajectory to remain within the target SoC band prior to critical hours. Gradients propagate through the dispatch layer back to the predictor, aligning forecasts with the operational objective, and improving the \emph{reachability} of SoC targets.

\subsubsection{Dispatch layer with stopping-time reward}
The dispatch layer with stopping-time reward is presented below:
\begin{subequations} \label{P3}
    \begin{align}
        f(r,\lambda, c,d,e,z,g) = \;& \sum_{t=1}^{T} \lambda_{t} P (d_t - c_t) + \sum_{t=1}^{T} r_t z_{t} \label{P3:objective_function} \\
        & + \rho ||e_T - e^{\rm target}||^2\nonumber\\
        \text{s.t.}\quad& z_{t=0} = 0, ~e_{t=0} = e_0 \label{P3:initial_states}\\
        \Big\{ & \eqref{P0:storage_capacity} - \eqref{P0:battery_dynamics}, \nonumber\\
        & \eqref{P1:stop_monotonicity}- \eqref{P1:charge_discharge_idle_limits}, \nonumber\\
        & z_{t},~c_{t},~d_{t},~g_{t}\in [0,1] \Big\} \qquad \forall~t\in~\mathcal{T}, \nonumber
    \end{align}
\end{subequations}
where $z_t \in [0,1]$ is a continuous variable that represents if the battery asset starts the reserve-preserving mode. $g_t$ corresponds to the battery idle action that works as a slack variable to delay the decision of entering into reserve-preserving mode. Eq. \eqref{P3:objective_function} is the objective function that includes the arbitrage reward, stopping reward factor $r_t$, and deviation penalties. The SoC target $e^{\rm target}$ is proposed as the midpoint of $[\ell^{\rm min}, \ell^{\rm max}]$, but it is assumed to be set by the system operator (e.g., through the day-ahead schedule or, experimentally, through the tolerance to critical-constraint violations). The sequence $\{r_t\}_{t=1}^{T}$ is chosen by the system operator capturing the system's risk awareness. 

\subsubsection{Uncertainty sets and conformal calibration}
Based on the work in \cite{yeh2024end}, we implement a partially-input-convex neural network uncertainty set for energy price uncertainty. Conformal calibration is used to calibrate this uncertainty set to solve the conditional robust arbitrage problem. First, we fit the point predictor $\hat{\lambda}_t$ on the training set. In a calibration split procedure, the residuals are formed between the realized prices $\lambda_t$ and the predicted centers $\hat{\lambda}_t$, and select hourly quantiles $q_{1-\delta}$ to construct bounds $[\hat{\lambda}_t-q_{1-\delta},\,\hat{\lambda}_t+q_{1-\delta}]$ that achieve finite-sample coverage $1-\delta$ for future price realizations.

Consider $x$ as an exogenous feature for energy prices $\lambda = \lambda(x)$. Then, the price predictor outputs a center $\hat\lambda_t(x)$ and an uncertainty set $\mathcal{U}_\theta(x)$, where $\theta$ represents the parameters of the predictive model. The dispatch layer aims to maximize the conditional robust problem under the worst-case scenario of \eqref{P3:objective_function} over the estimated and calibrated $\lambda \in \mathcal{U}_\theta(x)$.

\begin{figure}[ht]
    \centering
    \includegraphics[width=1.02\linewidth, trim={4.5cm 16.4cm 6.3cm 9.6cm},clip]{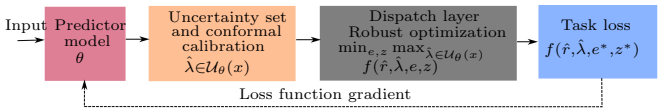}
    \caption{End-to-end framework considered for the energy arbitrage with stopping-time reward.}
    \label{fig:enter-label1}
    \vspace{-1mm}
\end{figure}

Figure \ref{fig:enter-label1} shows the framework pipeline. Given input training samples $\{(x_i,\lambda_i)\}_{i\in N}$ and a calibration split process, each iteration proceeds as follows: (i) estimate price predictor parameters $\theta$; (ii) compute residuals on the mini-batch calibration subset $N$, set the quantile threshold $q$ to the $(1-\delta)$ empirical quantile, and construct $\mathcal{U}_\theta(x)$; (iii) solve the robust dispatch problem for each $x_i$ to obtain $y\!=\!(c,d,g,e,z)$ and the resulting per-sample task loss $\ell_i(\theta) = -\;f\!\big(\hat{r},\hat{\boldsymbol{\lambda}}_i(\theta),y^*\big)$; (iv) compute the task loss; and (v) back-propagate $\nabla_\theta \ell_i$ through the optimization layer using the implicit-function gradient and update $\theta$.

This pipeline jointly learns forecasts and uncertainty widths that optimize arbitrage profit while meeting reliability, and the stopping reward shapes the learned policy so that SoC trajectories remain within the SoC target band during critical hours. The method retains convexity, yields exact constraint satisfaction, and provides finite-sample coverage for price uncertainty using conformal calibration, while the stopping-time design incorporates operator risk preferences directly in the objective.

\section{Case Study}
In this section, we compare the performance of the SAA, DQN, and E2E models in terms of the expected profit, stopping-time decisions, and the total reward for a battery asset performing energy arbitrage. Given a particular realization of prices, $\{\lambda_t\}_{t=1}^T$, the battery owner seeks to optimize their objective that combines arbitrage profit, penalties for violating a minimum state-of-charge (SoC) target range, and a per-period reward accumulated after stopping.

We use six years of PJM real-world grid data, $\{(x_i, \lambda_i)\}_{i=1}^{N}$, similarly to  \cite{donti2017task}, which includes temperature and energy price information. The time horizon is set to $T = 24$, and the set of critical hours is a singleton, $\mathcal{T}^{\rm target} = \{24\}$. All models are evaluated in a common set of 438 out-of-sample scenarios to enable paired comparisons and robust statistical analysis. The charging/discharging and self-discharge efficiencies are $\eta_c = \eta_d = 0.9$ and $\hat{\eta} = 0.995$, respectively. The battery limits are set to $e^{\min}=0$ MWh and $e^{\max}=10$ MWh. The maximum charge and discharge rate is $P=5$ MW. The initial SoC $e_{t=0}\!=\!\frac{e^{\rm max}}{2}$ MWh. Finally, the target band $\mathcal{E}^{\ell} = [5,7]$ MWh and $e^{\rm target} = 6$ MWh. We compare models performance and battery decisions taken under different constant values for the stopping-reward series $\{r_t\}_{t=1}^{24} = c$ with $c = \{0,\ldots,15\} \frac{\$}{\rm hr}$.

The SAA model shows higher profits due to the lack of penalty terms. Considering, $c=0\frac{\$}{\rm hr}$ the model profit is $97.6\$$ with $45\%$ deviation. These values remain constant until the reward value reaches $c\in\{3,\ldots,7\}\frac{\$}{\rm hr}$, where the profit increases on average by $9\%$ for each increase in $c$, the standard deviation is reduced to $25\%$, and the stopping-time is $\tau = 13$ hr. For values $c\in\{8,\ldots,11\}\frac{\$}{\rm hr}$, the arbitrage profit decreases $41\%$ relative to $c=0\frac{\$}{\rm hr}$ and remains constant for all values of $c$ in that range. The stopping-reward increases linearly since the stopping-time is maintained at $\tau = 5$ hr. For $c > 11\frac{\$}{\rm hr}$, the stopping-time changes to $1$ hr, with all profit coming from the stopping reward. Thus, by construction, the final SoC remains at $5$ MWh, and the stopping-time pattern changes in a stepwise manner with respect to $c$, with discrete shifts observed at $c\in\{3,7,11\}\frac{\$}{\rm hr}$.

For the test samples, with $c=0\frac{\$}{hr}$, the DQN model yields a profit of $\$161$ with a standard deviation of $94\%$. For the reward range $c\in\{1,\ldots,11\}\frac{\$}{\rm hr}$, the total profit and the standard deviation remain around the same, with average stopping-time as $\tau=23$ hr. For $c\in\{11,\ldots,13\}\frac{\$}{\rm hr}$, the average stopping-time changes to $\tau=2$ hr, the profit increases by $41\%$, and the standard deviation increases to $128\%$. Finally, for $c>12\frac{\$}{\rm hr}$, the stopping-time is $\tau=1$ hr, the profit increases by $110\%$ relative to $c\!=\!0\frac{\$}{hr}$, and the standard deviation is reduced to $0.1\%$.

Under the DQN model, large penalty terms can dominate the Q-targets, so the greedy policy under-weights the stopping-time reward even when the terminal payoff is relatively large. Action values are also affected by the relatively large terminal penalty (set to $\rho = 10$), and changing the stopping reward effectively defines a different feasible space, which inhibits consistent comparison of results under different reward specifications. In this case, the learned stopping-times are, on average, consistent across runs but there is some variation driven by the valuation of successor states $Q_{\theta}(s_{t+1})$, e.g., solutions with $c=6\frac{\$}{\rm hr}$. Reward variance increases and profits become not directly comparable unless penalties and rewards are normalized to a common scale. For example, if the target penalty increases to $\rho = 1000$, then for $c=5\frac{\$}{\rm hr}$ the stopping-time changes to $\tau=1$ hr, while for $c=10\frac{\$}{\rm hr}$ it changes to $\tau=2$ hr, resulting in negative task profits.

For the E2E framework, with $c=0\frac{\$}{\rm hr}$ the expected profit is $\$0$ for the test scenarios, which is lower than in the SAA case due to its robust design. With $c\in\{1,\ldots,6\}\frac{\$}{\rm hr}$, the expected profit increases to $\$130$ with a standard deviation of $3\%$ and increases nearly linearly, while the average stopping-time is $\tau = 16$ hr. Increasing to $c\in\{7,\ldots,15\}\frac{\$}{\rm hr}$, the average profit increases by about $9\%$ as $c$ increases, with a standard deviation of $1\%$, with stopping-time $\tau = 18$ hr. Across all simulations, the marginal coverage, i.e., the percentage of price realizations considered by the robust optimization problem, remains above $94\%$. Thus, the E2E model delivers lower variability in profit and a more consistent stopping-time reward and decision, while maintaining high coverage across price realizations. In practice, this robustness trades off average performance for stability, concentrating outcomes and reducing sensitivity to price perturbations.

\begin{figure}[http]
  \centering
  \includegraphics[width=\linewidth]{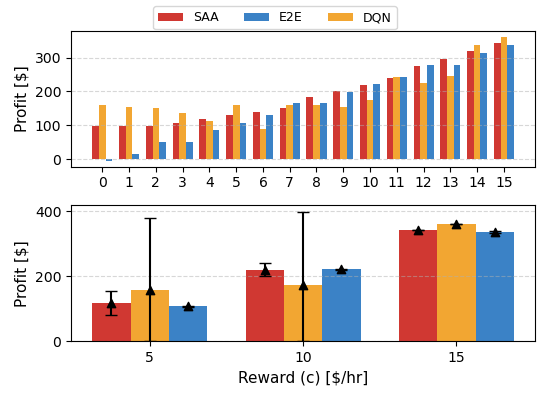}
  \caption{(top) Total profit between models for $c = \{0,\ldots,15\} \frac{\$}{\rm hr}$, (bottom) variance for $c= \{5,10,15\}\frac{\$}{\rm hr}$.}
  \label{fig:enter-label3}
  \vspace{-2mm}
\end{figure}
\begin{figure}[http]
  \centering
  \includegraphics[width=\linewidth]{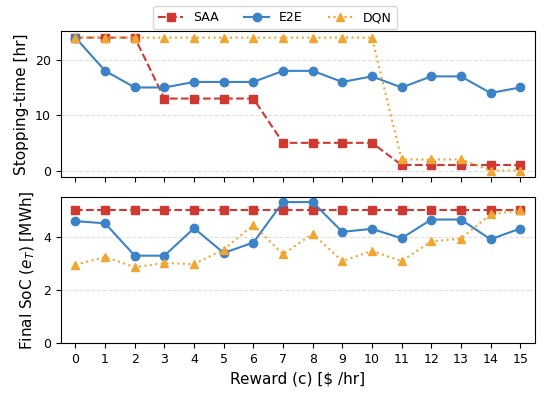}
  \caption{(top) Stopping-time, (bottom) final SoC between models for $c = \{0,\ldots,15\} \frac{\$}{\rm hr}$.}
  \label{fig:enter-label4}
  \vspace{-3mm}
\end{figure}
Figure \ref{fig:enter-label3} shows the total profit under different reward values. The top sub-figure shows that DQN model over-performs the rest of the models due to its online optimization structure. In addition, the E2E model profits approach the rest of the models when $c>7\frac{\$}{\rm hr}$. For E2E, with $c\in\{3,\ldots,7\}\,\frac{\$}{\mathrm{hr}}$, the stopping reward represents $31$-$51\%$ of the total profit, for $c\in\{8,\ldots,12\}\,\frac{\$}{\mathrm{hr}}$, it accounts for $83$-$87\%$, and it reaches $100\%$ for $c>12\,\frac{\$}{\mathrm{hr}}$. In contrast, for DQN the stopping reward begins to represent a substantial percentage of the total profit only for $c\in\{11,\ldots,15\}\,\frac{\$}{\mathrm{hr}}$, accounting for $89$-$100\%$. The bottom sub-figure shows that the DQN model has high variance, sometimes producing near negative values due myopic sensitivity to the penalties until the end of the horizon. In comparison, the E2E model achieves profits close to those of the other models, with a small standard deviation when $c>7\frac{\$}{\rm hr}$. Figure~\ref{fig:enter-label4} shows the stopping-time decision and the final SoC, where the top sub-figure shows how the stopping-time tends to decrease as $c$ increases. The DQN model exhibits a pronounced stepwise behavior, while the E2E model remains stable because the stopping decisions are relaxed and the final stopping-time is estimated using the net stopping reward. The lower sub-figure shows that both models tend to undershoot the safe SoC band, resulting in a final SoC of about 3–4 MWh.

Table \ref{tab:model-comp} summarizes the optimal profits, learned stopping-time, and the average stopping-time rewards for $c\!=\!\{5,10,15\}\frac{\$}{\rm hr}$. Across the formulations, the chosen stopping-times shift due to the model's conservativeness toward price uncertainty. We observe that compared to the DQN model, the E2E consistently yields a similar average task profit, for $c>7\frac{\$}{\mathrm{hr}}$, and reduces variability. Table \ref{tabletable} describes the presented models in terms of their objectives and reliability parameters.

\begin{table}[http]
  \centering
  \caption{Optimal profit, learned stopping-time, and average stopping–reward by model and reward level.}
  \label{tab:model-comp}
  \small
  \begin{tabular}{lcccc}
    \toprule
    \multirow{2}{*}{Model} &\textbf{Reward} & \textbf{Optimal} & \textbf{Stopping-time} & \textbf{Stopping} \\
     & \textbf{$c$ [\$/hr]}& \textbf{profit [\$]} & \textbf{$\tau$ [hr]} & \textbf{reward [\$]} \\
    \midrule
    \multirow{3}{*}{DQN} & 5  & 158.8 & 23 & 0.7 \\
                         & 10 & 174.5 & 23 & 6.1 \\
                         & 15 & 360.0 & 1 & 360.0 \\
    \midrule
    \multirow{3}{*}{E2E} & 5  & 107.65 & 16 & 42.37 \\
                         & 10 & 221.27 & 17 & 77.54\\
                         & 15 & 335.56 & 15 & 141.29\\
    \bottomrule
  \end{tabular}
  \vspace{-6pt}
\end{table}

\begin{table}[http]
\centering
\caption{Modeling alternatives for battery arbitrage with reliability targets.}
\label{tab:model-comp2}
\small
\begin{tabular}{lcccc}
\toprule
\textbf{Model} & \textbf{Objective} & \textbf{Reliability} & \textbf{Pros/Cons} \\
\midrule
SAA & \eqref{P2:lower_bound_stopping}-\eqref{P2:minimum_scenarios} & Empirical $(1-\epsilon)$ & Comp. expensive \\
DQN & \eqref{dqn_state}-\eqref{dqn_reserve_preserving} & Tunable via $\hat{r}_t,\rho,\gamma$ & Model-free \\
E2E  & \eqref{P2} & Tunable via $\hat{r}_t,\rho$ & Task-driven\\
\bottomrule
\end{tabular}
\vspace{-6pt}
\label{tabletable}
\end{table}

Note that the DQN and E2E models show different profit magnitudes because they incur extra penalties for violating the SoC range and for not satisfying feasibility constraints. Since historical prices include hard-to-predict spikes, the performance of the E2E and DQN models is relatively variable and depends on how well model parameters are tuned \cite{donti2017task}.

\section{Conclusion and Future Work}
This paper proposes using the stopping-time criterion as an attribute of reachability analysis, used to cope with \emph{critical hours}. Incorporating this criterion allows for steering storage assets toward feasible trajectories and improving reachability to the SoC range. Among the three models compared, the E2E model yields the least variability in decisions, while also producing stable stopping-times. By contrast, the DQN method frequently yields higher profit magnitudes, higher variance, and occasional feasibility violations. 

Future work will extend the framework to predictive control of the battery considering a stopping reward, and extend the stopping-reward design to quantify the system operator's risk and  study how other time-series dynamics can steer battery assets to deliver resilience attributes to the system.

\section*{Acknowledgement}
This work was supported in part by the US DOE ARPA-e under Grant DE-AR0001300 and in part by the US National Science Foundation under Grant OISE 2330450.

\bibliographystyle{IEEEtran}
\bibliography{bib}

@article{maov_battery2013,
author = {Ortega-Vazquez, Miguel A.},
title = {Optimal scheduling of electric vehicle charging and vehicle-to-grid services at household level including battery degradation and price uncertainty},
journal = {IET Generation, Transmission \& Distribution},
volume = {8},
number = {6},
pages = {1007-1016},
doi = {https://doi.org/10.1049/iet-gtd.2013.0624},
url = {https://ietresearch.onlinelibrary.wiley.com/doi/abs/10.1049/iet-gtd.2013.0624},
eprint = {https://ietresearch.onlinelibrary.wiley.com/doi/pdf/10.1049/iet-gtd.2013.0624},
year = {2014}
}

@article{yang2020projection,
  title={Projection-based constrained policy optimization},
  author={Yang, Tsung-Yen and Rosca, Justinian and Narasimhan, Karthik and Ramadge, Peter J},
  journal={arXiv preprint arXiv:2010.03152},
  year={2020}
}

@article{ding2020natural,
  title={Natural policy gradient primal-dual method for constrained markov decision processes},
  author={Ding, Dongsheng and Zhang, Kaiqing and Basar, Tamer and Jovanovic, Mihailo},
  journal={Advances in Neural Information Processing Systems},
  volume={33},
  pages={8378--8390},
  year={2020}
}

@techreport{CAISO2024,
    author = {CAISO},
    title = {2023 Special Report on Battery Storage},
    institution = {CAISO},
    year = {2024}
}

@article{lechowicz2023online,
  title={The online pause and resume problem: Optimal algorithms and an application to carbon-aware load shifting},
  author={Lechowicz, Adam and Christianson, Nicolas and Zuo, Jinhang and Bashir, Noman and Hajiesmaili, Mohammad and Wierman, Adam and Shenoy, Prashant},
  journal={Proceedings of the ACM on Measurement and Analysis of Computing Systems},
  volume={7},
  number={3},
  pages={1--32},
  year={2023},
  publisher={ACM New York, NY, USA}
}

@article{yeh2024end,
  title={End-to-End Conformal Calibration for Optimization Under Uncertainty},
  author={Yeh, Christopher and Christianson, Nicolas and Wu, Alan and Wierman, Adam and Yue, Yisong},
  journal={arXiv preprint arXiv:2409.20534},
  year={2024}
}

@article{zheng2022arbitraging,
  title={Arbitraging variable efficiency energy storage using analytical stochastic dynamic programming},
  author={Zheng, Ningkun and Jaworski, Joshua and Xu, Bolun},
  journal={IEEE Transactions on Power Systems},
  volume={37},
  number={6},
  pages={4785--4795},
  year={2022},
  publisher={IEEE}
}

@inproceedings{lee2024online,
  title={Online Search with Predictions: Pareto-optimal Algorithm and its Applications in Energy Markets},
  author={Lee, Russell and Sun, Bo and Hajiesmaili, Mohammad and Lui, John CS},
  booktitle={Proceedings of the 15th ACM International Conference on Future and Sustainable Energy Systems},
  pages={386--407},
  year={2024}
}

@article{lorenz2009optimal,
  title={Optimal algorithms for k-search with application in option pricing},
  author={Lorenz, Julian and Panagiotou, Konstantinos and Steger, Angelika},
  journal={Algorithmica},
  volume={55},
  number={2},
  pages={311--328},
  year={2009},
  publisher={Springer}
}

@inproceedings{singhal2020pricing,
  title={Pricing impacts of state of charge management options for electric storage resources},
  author={Singhal, Nikita G and Ela, Erik G},
  booktitle={2020 IEEE Power \& Energy Society General Meeting (PESGM)},
  pages={1--6},
  year={2020},
  organization={IEEE}
}

@article{paternain2022safe,
  title={Safe policies for reinforcement learning via primal-dual methods},
  author={Paternain, Santiago and Calvo-Fullana, Miguel and Chamon, Luiz FO and Ribeiro, Alejandro},
  journal={IEEE Transactions on Automatic Control},
  volume={68},
  number={3},
  pages={1321--1336},
  year={2022},
  publisher={IEEE}
}

@inproceedings{achiam2017constrained,
  title={Constrained policy optimization},
  author={Achiam, Joshua and Held, David and Tamar, Aviv and Abbeel, Pieter},
  booktitle={International conference on machine learning},
  pages={22--31},
  year={2017},
  organization={PMLR}
}

@inproceedings{xu2020operational,
  title={Operational valuation of energy storage under multi-stage price uncertainties},
  author={Xu, Bolun and Korp{\aa}s, Magnus and Botterud, Audun},
  booktitle={2020 59th IEEE Conference on Decision and Control (CDC)},
  pages={55--60},
  year={2020},
  organization={IEEE}
}

@article{zhang2021arbitrage,
  title={Arbitrage analysis for different energy storage technologies and strategies},
  author={Zhang, Xinjing and Qin, Chao Chris and Loth, Eric and Xu, Yujie and Zhou, Xuezhi and Chen, Haisheng},
  journal={Energy Reports},
  volume={7},
  pages={8198--8206},
  year={2021},
  publisher={Elsevier}
}

@article{zhang2023day,
  title={Day-ahead optimization dispatch strategy for large-scale battery energy storage considering multiple regulation and prediction failures},
  author={Zhang, Mingze and Li, Weidong and Yu, Samson Shenglong and Wen, Kerui and Muyeen, SM},
  journal={Energy},
  volume={270},
  pages={126945},
  year={2023},
  publisher={Elsevier}
}

@article{qi2025locational,
  title={Locational energy storage bid bounds for facilitating social welfare convergence},
  author={Qi, Ning and Xu, Bolun},
  journal={IEEE Transactions on Energy Markets, Policy and Regulation},
  year={2025},
  publisher={IEEE}
}

@article{dvorkin2016ensuring,
  title={Ensuring profitability of energy storage},
  author={Dvorkin, Yury and Fernandez-Blanco, Ricardo and Kirschen, Daniel S and Pand{\v{z}}i{\'c}, Hrvoje and Watson, Jean-Paul and Silva-Monroy, Cesar A},
  journal={IEEE Transactions on Power Systems},
  volume={32},
  number={1},
  pages={611--623},
  year={2016},
  publisher={IEEE}
}

@article{qi2023chance,
  title={Chance-constrained generic energy storage operations under decision-dependent uncertainty},
  author={Qi, Ning and Pinson, Pierre and Almassalkhi, Mads R and Cheng, Lin and Zhuang, Yingrui},
  journal={IEEE Transactions on Sustainable Energy},
  volume={14},
  number={4},
  pages={2234--2248},
  year={2023},
  publisher={IEEE}
}

@article{harsha2014optimal,
  title={Optimal management and sizing of energy storage under dynamic pricing for the efficient integration of renewable energy},
  author={Harsha, Pavithra and Dahleh, Munther},
  journal={IEEE Transactions on Power Systems},
  volume={30},
  number={3},
  pages={1164--1181},
  year={2014},
  publisher={IEEE}
}

@article{shapiro2012time,
  title={Time consistency of dynamic risk measures},
  author={Shapiro, Alexander},
  journal={Operations Research Letters},
  volume={40},
  number={6},
  pages={436--439},
  year={2012},
  publisher={Elsevier}
}

@article{pichler2022risk,
  title={Risk-averse stochastic programming: Time consistency and optimal stopping},
  author={Pichler, Alois and Liu, Rui Peng and Shapiro, Alexander},
  journal={Operations Research},
  volume={70},
  number={4},
  pages={2439--2455},
  year={2022},
  publisher={INFORMS}
}

@article{bielecki2025time,
  title={Time consistency of dynamic risk measures and dynamic performance measures generated by distortion functions},
  author={Bielecki, Tomasz R and Cialenco, Igor and Liu, Hao},
  journal={Stochastic Models},
  volume={41},
  number={2},
  pages={180--207},
  year={2025},
  publisher={Taylor \& Francis}
}

@article{sioshansi2009estimating,
  title={Estimating the value of electricity storage in PJM: Arbitrage and some welfare effects},
  author={Sioshansi, Ramteen and Denholm, Paul and Jenkin, Thomas and Weiss, Jurgen},
  journal={Energy economics},
  volume={31},
  number={2},
  pages={269--277},
  year={2009},
  publisher={Elsevier}
}

@article{cao2020deep,
  title={Deep reinforcement learning-based energy storage arbitrage with accurate lithium-ion battery degradation model},
  author={Cao, Jun and Harrold, Dan and Fan, Zhong and Morstyn, Thomas and Healey, David and Li, Kang},
  journal={IEEE Transactions on Smart Grid},
  volume={11},
  number={5},
  pages={4513--4521},
  year={2020},
  publisher={IEEE}
}

@article{degroot1968some,
  title={Some problems of optimal stopping},
  author={DeGroot, Morris H},
  journal={Journal of the Royal Statistical Society Series B: Statistical Methodology},
  volume={30},
  number={1},
  pages={108--122},
  year={1968},
  publisher={Oxford University Press}
}

@inproceedings{wang2018energy,
  title={Energy storage arbitrage in real-time markets via reinforcement learning},
  author={Wang, Hao and Zhang, Baosen},
  booktitle={2018 IEEE Power \& Energy Society General Meeting (PESGM)},
  pages={1--5},
  year={2018},
  organization={IEEE}
}

@article{krishnamurthy2017energy,
  title={Energy storage arbitrage under day-ahead and real-time price uncertainty},
  author={Krishnamurthy, Dheepak and Uckun, Canan and Zhou, Zhi and Thimmapuram, Prakash R and Botterud, Audun},
  journal={IEEE Transactions on Power Systems},
  volume={33},
  number={1},
  pages={84--93},
  year={2017},
  publisher={IEEE}
}

@article{babaioff2008online,
  title={Online auctions and generalized secretary problems},
  author={Babaioff, Moshe and Immorlica, Nicole and Kempe, David and Kleinberg, Robert},
  journal={ACM SIGecom Exchanges},
  volume={7},
  number={2},
  pages={1--11},
  year={2008},
  publisher={ACM New York, NY, USA}
}

@article{cheng2016co,
  title={Co-optimizing battery storage for the frequency regulation and energy arbitrage using multi-scale dynamic programming},
  author={Cheng, Bolong and Powell, Warren B},
  journal={IEEE Transactions on Smart Grid},
  volume={9},
  number={3},
  pages={1997--2005},
  year={2016},
  publisher={IEEE}
}

@article{donti2017task,
  title={Task-based end-to-end model learning in stochastic optimization},
  author={Donti, Priya and Amos, Brandon and Kolter, J Zico},
  journal={Advances in neural information processing systems},
  volume={30},
  year={2017}
}

@article{li2020online,
  title={Online optimization with predictions and switching costs: Fast algorithms and the fundamental limit},
  author={Li, Yingying and Qu, Guannan and Li, Na},
  journal={IEEE Transactions on Automatic Control},
  volume={66},
  number={10},
  pages={4761--4768},
  year={2020},
  publisher={IEEE}
}

@article{sang2022electricity,
  title={Electricity price prediction for energy storage system arbitrage: A decision-focused approach},
  author={Sang, Linwei and Xu, Yinliang and Long, Huan and Hu, Qinran and Sun, Hongbin},
  journal={IEEE Transactions on Smart Grid},
  volume={13},
  number={4},
  pages={2822--2832},
  year={2022},
  publisher={IEEE}
}

@article{zhang2011online,
  title={Online algorithms for the general k-search problem},
  author={Zhang, Wenming and Xu, Yinfeng and Zheng, Feifeng and Liu, Ming},
  journal={Information processing letters},
  volume={111},
  number={14},
  pages={678--682},
  year={2011},
  publisher={Elsevier}
}

@article{bienstock2024risk,
  title={Risk-Aware Security-Constrained Unit Commitment: Taming the Curse of Real-Time Volatility and Consumer Exposure},
  author={Bienstock, Daniel and Dvorkin, Yury and Guo, Cheng and Mieth, Robert and Wang, Jiayi},
  journal={IEEE Transactions on Energy Markets, Policy and Regulation},
  year={2024},
  publisher={IEEE}
}

@article{dietrich2025data,
  title={Data-Driven Reachability with Scenario Optimization and the Holdout Method},
  author={Dietrich, Elizabeth and Devonport, Rosalyn and Tu, Stephen and Arcak, Murat},
  journal={arXiv preprint arXiv:2504.06541},
  year={2025}
}

@article{jafarpour2025probabilistic,
  title={Probabilistic reachability analysis of stochastic control systems},
  author={Jafarpour, Saber and Liu, Zishun and Chen, Yongxin},
  journal={IEEE Transactions on Automatic Control},
  year={2025},
  publisher={IEEE}
}

@article{zhao2024conformal,
  title={Conformal predictive programming for chance constrained optimization},
  author={Zhao, Yiqi and Yu, Xinyi and Sesia, Matteo and Deshmukh, Jyotirmoy V and Lindemann, Lars},
  journal={arXiv preprint arXiv:2402.07407},
  year={2024}
}

@inproceedings{lew2021sampling,
  title={Sampling-based reachability analysis: A random set theory approach with adversarial sampling},
  author={Lew, Thomas and Pavone, Marco},
  booktitle={Conference on robot learning},
  pages={2055--2070},
  year={2021},
  organization={PMLR}
}

@phdthesis{althoff2010reachability,
  title={Reachability analysis and its application to the safety assessment of autonomous cars},
  author={Althoff, Matthias},
  year={2010},
  school={Technische Universit{\"a}t M{\"u}nchen}
}

@article{althoff2021set,
  title={Set propagation techniques for reachability analysis},
  author={Althoff, Matthias and Frehse, Goran and Girard, Antoine},
  journal={Annual Review of Control, Robotics, and Autonomous Systems},
  volume={4},
  number={1},
  pages={369--395},
  year={2021},
  publisher={Annual Reviews}
}

@article{wang2023learning,
  title={Learning decision-focused uncertainty sets in robust optimization},
  author={Wang, Irina and Becker, Cole and Van Parys, Bart and Stellato, Bartolomeo},
  journal={arXiv preprint arXiv:2305.19225},
  year={2023}
}

@inproceedings{castellano2020learning,
  title={Learning the operation of energy storage systems from real trajectories of demand and renewables},
  author={Castellano, Agustin and Bazerque, Juan Andr{\'e}s},
  booktitle={2020 IEEE Power \& Energy Society Innovative Smart Grid Technologies Conference (ISGT)},
  pages={1--5},
  year={2020},
  organization={IEEE}
}

@techreport{PUCT2024,
    author = {PUCT},
    title = {{1186NPRR-36} PUCT REPORT 041124},
    institution = {PUCT},
    year = {2024}
}

@misc{PGE_SAVE_VPP_2025,
  author       = {{Pacific Gas and Electric Company}},
  title        = {{PG\&E} Launches Seasonal Aggregation of Versatile Energy ({SAVE}) Virtual Power Plant Program},
  organization = {PG\&E Corporation},
  address      = {Oakland, CA},
  year         = {2025},
  month        = mar,
  day          = {24},
  urldate      = {2025-10-13}
}

@article{schleifer2025exploring,
  title={Exploring the Future Energy Value of Long-Duration Energy Storage},
  author={Schleifer, Anna H and Cohen, Stuart M and Cole, Wesley and Denholm, Paul and Blair, Nate},
  journal={Energies},
  volume={18},
  number={7},
  pages={1751},
  year={2025},
  publisher={MDPI}
}

@inproceedings{porteiro2018towards,
  title={Towards multi-timescale energy provisioning using stochastic dual dynamic programming},
  author={Porteiro, Rodrigo and Ferragut, Andres and Paganini, Fernando},
  booktitle={2018 IEEE 9th Power, Instrumentation and Measurement Meeting (EPIM)},
  pages={1--6},
  year={2018},
  organization={IEEE}
}

@article{pereira1991multi,
  title={Multi-stage stochastic optimization applied to energy planning},
  author={Pereira, Mario VF and Pinto, Leontina MVG},
  journal={Mathematical programming},
  volume={52},
  number={1},
  pages={359--375},
  year={1991},
  publisher={Springer}
}

@article{garcia1989model,
  title={Model predictive control: Theory and practice—A survey},
  author={Garcia, Carlos E and Prett, David M and Morari, Manfred},
  journal={Automatica},
  volume={25},
  number={3},
  pages={335--348},
  year={1989},
  publisher={Elsevier}
}

@article{castellano2025data,
  title={Data-driven Acceleration of {MPC} with Guarantees},
  author={Castellano, Agustin and Pan, Shijie and Mallada, Enrique},
  journal={arXiv preprint arXiv:2511.13588},
  year={2025}
}

@inproceedings{castellano2023learning,
  title={Learning safety critics via a non-contractive binary bellman operator},
  author={Castellano, Agustin and Min, Hancheng and Bazerque, Juan Andr{\'e}s and Mallada, Enrique},
  booktitle={2023 57th Asilomar Conference on Signals, Systems, and Computers},
  pages={814--821},
  year={2023},
  organization={IEEE}
}

\end{document}